# Are Heptazine-Based Organic Light-Emitting Diode (OLED) Chromophores Thermally Activated Delayed Fluorescence (TADF) or Inverted Singlet-Triplet (IST) Systems?

Andrzej L. Sobolewski[1] and Wolfgang Domcke[2*]

[1] Institute of Physics, Polish Academy of Sciences, Warsaw, Poland

[2] Department of Chemistry, Technical University of Munich, Garching, Germany

* corresponding author; Email: domcke@ch.tum.de

**Abstract**

Two chromophores derived from heptazine, HAP-3MF and HAP-3TPA, were synthesized and tested as emitters in light-emitting diodes (OLEDs) by Adachi and coworkers. Both emitters were shown to exhibit quantum efficiencies which exceed the theoretical maximum of conventional fluorescent OLEDs. The enhanced emission efficiency was explained by the mechanism of thermally activated delayed fluorescence (TADF). In the present work, the electronic excitation energies and essential features of the topography of the excited-state potential-energy surfaces of HAP-3MF and HAP-3TPA have been investigated with a wave-function based ab initio method (ADC(2)). It is found that HAP-3MF is an inverted singlet-triplet (IST) system, that is, the energies of the $S_1$ and $T_1$ states are robustly inverted in violation of Hund's multiplicity rule. Notably, HAP-3MF presumably is the first IST emitter which was implemented in an OLED device. In HAP-3TPA, on the other hand, the vertical excitation energies of the $S_1$ and $T_1$ states are essentially degenerate. The excited states exhibit vibrational stabilization energies of similar magnitude along different relaxation coordinates, resulting in adiabatic excitation energies which also are nearly degenerate. HAP-3TPA is found to be a chromophore at the borderline of TADF and IST systems. The spectroscopic data reported by Adachi and coworkers for HAP-3MF and HAP-3TPA are analysed in the light of these computational results

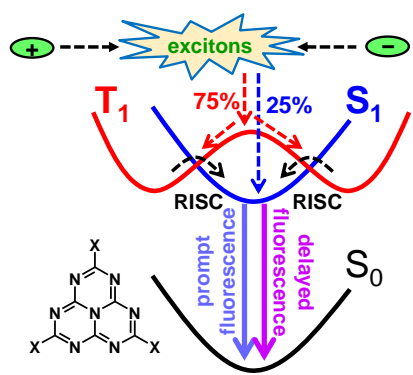

**TOC graphics**

The continuous improvement of the efficiency of organic light emitting diodes (OLEDs) by the optimization of the emitting chromophores is currently a very active field of research. The statistical recombination of injected electrons and holes in light emitting materials results in 75% triplet excitons and 25% singlet excitons. To harvest the non-fluorescent triplet excitons, the concepts of reverse intersystem crossing (RISC) and thermally activated delayed fluorescence (TADF) have been invoked.[1-5] If the singlet-triplet gap, $\Delta_{ST} = E_S - E_T$, is of the order of $k_B T$, where $k_B$ is the Boltzmann constant and $T$ is the temperature, the population of the lowest triplet state ($T_1$) can be converted via thermal fluctuations and spin-orbit (SO) coupling to the higher-lying lowest singlet state ($S_1$) from which fluorescence can be emitted. It has been reported that with certain TADF emitters nearly 100% internal quantum efficiency of luminescence can be achieved.[2-5]

For $S_1$ and $T_1$ states arising from the same orbital configuration, $\Delta_{ST}$ is determined by the exchange integral. The exchange integral can become small when the highest occupied molecular orbital (HOMO) and the lowest unoccupied molecular orbital (LUMO), which usually are involved in the $S_0$-$S_1$ and $S_0$-$T_1$ transitions, are spatially separated, which can be realized for charge-transfer (CT) excitations in extended donor-acceptor (DA) systems.[5-9] Numerous systems involving aromatic donors and acceptors have been synthesized and evidence of the existence of singlet-triplet gaps of the order of 0.1 – 0.5 eV has been reported.[6,9-13] The disadvantage of this construction principle is the inevitably low oscillator strength of the $S_1$-$S_0$ transition in the case of non-overlapping HOMO and LUMO orbitals. Moreover, the SO coupling between singlet and triplet states of the same orbital configuration is very small according to the El-Sayed rules.[14] Another disadvantage of CT states as emitting states is the inherently broadly distributed fluorescence and therefore low color purity.[11,15] A more general model involving locally excited states as well as CT states was proposed by de Silva et al.[16] This model can explain the coexistence of an appreciable $T_1 \rightarrow S_1$ RISC rate with a nonvanishing fluorescence rate of the $S_1$ state.

Recently, an alternative concept has been promoted by Hatakeyama and coworkers, which builds on short-range internal CT states in planar boron/nitrogen heterocycles.[17-21] In these systems, HOMO and LUMO can be preferentially localized on different atoms which also reduces the magnitude of the exchange integral, resulting in a small $\Delta_{ST}$. Singlet-triplet gaps of the order of 0.2 – 0.4 eV have been determined.[15] Due to the rigidity of these chromophores, very narrow luminescence spectra can be realized, resulting in high color purity.[18,19,21]

A relatively new class of organic chromophores are inverted singlet-triplet (IST) emitters, which exhibit a negative $\Delta_{ST}$, that is, the energy minimum of the $S_1$ state is below the energy minimum of the $T_1$ state in violation of Hund's multiplicity rule. In IST emitters, triplet excitons can relax to singlet excitons by conventional SO-induced intersystem crossing (ISC). Thermal activation is not needed for achieving 100% internal quantum efficiency. In 2019, it was independently shown by de Silva[22] and Sobolewski, Domcke and coworkers[23] with wave-function based ab initio methods that the N-heterocycles cycl[3.3.3]azine and heptazine (1,3,5,6,9,9b-heptaazaphenalene) exhibit robustly negative singlet-triplet gaps. The nearly pure HOMO-LUMO character of the $S_1$ and $T_1$ states of azaphenalenes and spatially non-overlapping HOMO and LUMO result in an exceptionally small exchange integral (0.12 eV for heptazine (Hz)). The very compact structure of these orbitals favors spin polarization in the singlet state[24, 25] which lowers the energy of the $S_1$ state below the energy of the $T_1$ state.[23] Spin polarization is represented by double excitations in the restricted Hartree-Fock orbital basis. The lack of double excitations in time-dependent density functional theory (TDDFT) is the reason why the TDDFT method with commonly used functionals cannot reproduce the singlet-triplet inversion.[22]

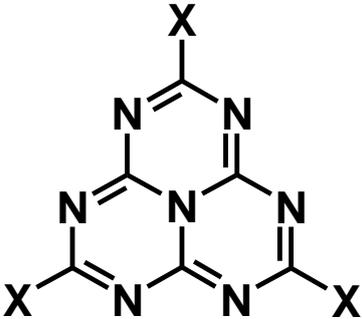

| Compound | X |
|---|---|
| (a) HAP-3MF | 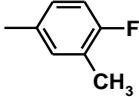 |
| (b) TAHz | 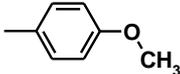 |
| (c) HAP-3TPA | 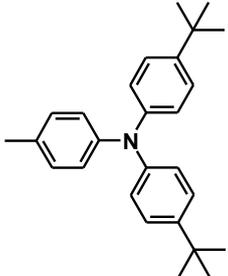 |
| (d) HAP-3TPA-t | 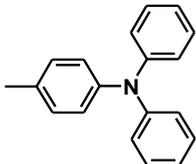 |

*Figure 1. Chemical structures of HAP-3MF (a), TAHz (b), HAP-3TPA (c), and truncated HAP-3TPA (HAP-3TPA-t)(d).*

Singlet-triplet inversion was experimentally confirmed for the Hz derivative 2,5,8-tris(4-methoxyphenyl)-1,3,5,6,9,9b-heptaazaphenalene or tri-anisole-heptazine (TAHz), which was synthesized and spectroscopically characterized by Rabe et al.[26] It was observed that the luminescence of TAHz is not quenched by oxygen and is insensitive to the presence of external heavy atoms.[21] Moreover, the theoretically predicted transient absorption of the triplet state could not be observed, while the predicted transient absorption of the exceptionally long-lived singlet state was readily detected.[23] More recently, Sancho-Garcia and coworkers confirmed the singlet-triplet inversion (and the failure of TDDFT to reproduce it) for nitrogen or boron containing triangulenes of various sizes.[27, 28] An extensive computational study of azaphenalenes of varying nitrogen content and with a large variety of aromatic pendants was performed by Pollice et al., focussing on the compatibility of a negative $\Delta_{ST}$ and a reasonably large oscillator strength of the $S_1$ state. It was predicted that IST chromophores with appreciable fluorescence rates exist.[29] It has also been shown that singlet-triplet inversion can be achieved in TADF-type chromophores using strong light-matter coupling in optical cavities.[30-32]

So far, two chromophores derived from Hz were synthesized and tested as emitters in OLEDs. In 2014, Adachi and coworkers tested an OLED which contained 2,5,8-tris(4-fluoro-3-methylphenyl)-1,3,4,6,7,9,9b-heptaazaphenalene (HAP-3MF), see Fig. 1a, as emitter.[33] In addition, an exiplex system was tested which contains HAP-3MF as electron-acceptor and 1,3-di(9H-carbazol-9-yl)benzene (mCP) as electron donor.[34] A high EQE of 11.3% was obtained with this exiplex. The exceptionally high quantum efficiency was explained by the TADF mechanism. It should be noted that HAP-3MF is isoelectronic with TAHz mentioned above, see Fig. 1b. In 2013, Adachi and coworkers synthesized 4,4',4''-(1,3,3a$^1$,4,6,7,9-heptaazaphenalene-2,5,8-triyl)tris(N,N-bis(4-(tert-butyl)phenyl)aniline (HAP-3TPA), which was conceived as a DA system with a heptazine core as electron-accepting unit and three substituted anilines as electron donors,[35] see Fig. 1c. An OLED containing HAP-3TPA as emitter was fabricated and an external quantum efficiency (EQE) of 17.5% was reported, which significantly exceeds the theoretical maximum of conventional fluorescent OLEDs.[35]

In the present work, we have explored the vertical and adiabatic excitation energies of HAP-3MF and HAP-3TPA with a wave-function-based ab initio method. For HAP-3TPA, hardware and software limitations required the cutting of the tert-butyl groups, that is, the electron donating pendants of HAP-3TPA were replaced by diphenylanilines, see Fig. 1d. This truncated compound is referred to as HAP-3TPA-t in the following. The excitation energies were computed with the ADC(2) (algebraic diagrammatic construction of second order) method,[36] which was also employed earlier for the investigation of the spectroscopy of TAHz.[23] The calculations provide clear evidence that HAP-3MF and TAHz are IST systems rather than TADF systems, that is, $\Delta_{ST}$ is negative both for the vertical excitation energies as well as for the adiabatic (minimum-to-minimum) excitation energies of the $S_1$ and $T_1$ states which are of A symmetry in the $C_3$ symmetry point group (*vide infra*). In HAP-3TPA-t, on the other hand, the situation is more complex. The large aromatic substituents of this chromophore lower the energy of a triplet state of E symmetry ($T_2$ in Hz) below the triplet state of A symmetry ($T_1$ in Hz) which arises from the HOMO-LUMO excitation. We find that the vertical excitation energy of the $^3$E state is degenerate with the vertical excitation energy of the $^1$A ($S_1$) state. The $^3$E state exhibits a Jahn-Teller (JT) effect via a degenerate vibrational mode, which lowers the energy minimum of the lower component of the $^3$E state. The energy of the $^1$A state, on the other hand, is stabilized by a distortion along a nondegenerate vibrational mode by about the same amount. The calculations predict that the energy minimum of the $^3$E state is merely 0.01 eV below the energy minimum of the $^1$A state. We discuss the interpretation of the photoluminescence (PL) and electroluminescence (EL) data of Adachi and coworkers in the light of these computational results. So far, HAP-3MF and HAP-3TPA are the only experimentally characterized Hz-based emitters in OLEDs. The rich experimental data for these systems and their interpretation on the basis of accurate ab initio calculations may provide inspiration for the future design of superior OLED emitters.

The ground-state equilibrium geometries of HAP-3MF and HAP-3TPA-t were determined with the second-order Møller-Plesset (MP2) *ab initio* method.[37] It was confirmed by the computation of the Hessian that the stationary points are minima. Vertical excitation energies (VEEs) of singlet and triplet excited states were calculated with the ADC(2) method.[36, 38-40] ADC(2) is a computationally efficient single-reference propagator method which yields similar excitation energies as the simplified second-order coupled-cluster (CC2) method.[41] The accuracy of CC2 and ADC(2) for excitation energies of organic molecules has been extensively benchmarked in comparison with more accurate methods, such as CC3 and EOM-CCSD.[42-44] Mean absolute errors of ≈ 0.22 eV for low-lying singlet states and ≈ 0.12 eV for low-lying triplet states have been estimated for the ADC(2) method with standard basis sets.[42] This accuracy is sufficient for the purposes of the present computational study. Excited-state geometry optimizations were performed with the ADC(2) method and adiabatic excitation energies and vertical emission energies to the electronic ground state were computed. The correlation-consistent polarized valence-split double-ζ basis set (cc-pVDZ)[45] was employed with the exception of HAP-3TPA-t, for which the use of a more compact basis set (def-SV(P)) was necessary to compute excited-state properties and to make the ADC(2) excited-state geometry optimizations feasible. The excitation energies computed with the def-SV(P) and cc-pVDZ basis sets typically differ by less than 0.04 eV. All calculations were performed with the TURBOMOLE program package (V. 6.3.1)[46] making use of the resolution-of-the-identity (RI) approximation.[47]

*Table 1. Vertical excitation energies (ΔE), oscillator strengths (f), dipole moments (μ), and leading electronic configurations of the lowest excited states of HAP-3MF determined with the ADC(2)/cc-pVDZ method at the MP2/cc-pVDZ equilibrium geometry of the ground state.*

| State | ΔE/eV | f | μ/Debye | Electronic Configuration |
|---|---|---|---|---|
| $S_0$ | 0.0 | - | 0.0 | (106a')$^2$(22a")$^2$ |
| $^1$A'(ππ*) | 2.58 | 0.0 | 0.0 | 0.75(20a"-23a")-0.48(19a"-23a") |
| $^3$A'(ππ*) | 2.82 | - | 0.0 | 0.77(20a"-23a")-0.48(19a"-23a") |
| $^3$E'(ππ*) | 3.48 | - | ±3.68 | 0.67(21a"-23a")/0.67(22a"-23a") |
| $^3$A"(nπ*) | 3.61 | - | 0.0 | 0.94(104a'-23a") |
| $^1$A"(nπ*) | 3.62 | 0.0 | 0.0 | 0.94(104a'-23a") |
| $^3$E"(nπ*) | 3.72 | - |  | 0.91(105a'-23a")/0.90(106a'-23a") |
| $^1$E"(nπ*) | 3.74 | 0.0 |  | 0.92(105a'-23a")/0.92(106a'-23a") |
| $^1$E'(ππ*) | 3.93 | 2.342 | ±6.62 | 0.78(21a"-23a")/0.78(22a"-23a") |

HAP-3MF exhibits $C_{3h}$ symmetry in the electronic ground state. For technical reasons, the ADC(2) computations were performed in $C_s$ symmetry. The VEEs of the four lowest singlet states and the four lowest triplet states of HAP-3MF calculated with the ADC(2) method are listed in Table 1. Oscillator strengths (for singlet states) and dipole moments (for degenerate states) are also given. In the singlet manifold, a dark ππ* state ($^1$A') is found at 2.58 eV, which is a characteristic feature of the excitation spectrum of the Hz chromophore.[23] The next higher singlet states are two $^1n\pi^*$ states at 3.62 eV ($^1$A″) and 3.74 eV ($^1$E″). The lowest bright state is the $^1$E′($\pi\pi^*$) state with a vertical excitation energy of 3.93 eV and with a high oscillator strength of 2.34. The calculated vertical excitation energy of the $^1$E′($\pi\pi^*$) state is in very good agreement with the observed absorption maximum of 3.80 eV.[33] The dark $^1n\pi^*$ states borrow intensity from the nearby bright $^1\pi\pi^*$ state and show up as weak peaks in the UV-vis absorption spectrum.[33]

The lowest triplet state ($^3$A′) is located 0.24 eV above the $^1$A′ state, that is, $\Delta_{ST}$ is negative and HAP-3MF clearly is an IST system. The $^3$E′ state at 3.48 eV is the triplet partner of the $^1$E′ state at 3.93 eV. For these higher states, $\Delta_{ST}$ is positive. As expected, the VEEs of HAT-3MF are very similar to those of isoelectronic TAHz which were reported previously.[23]

The VEEs of HAP-3MF computed with the widely used time-dependent DFT (TDDFT) method with the B3LYP functional are listed in Table S1 in the supporting information. The low-lying $S_1$ and $T_1$ states of $\pi\pi^*$ character are correctly predicted, but TDDFT is unable to predict the inversion of the energies of $S_1$ and $T_1$ states due to the lack of double excitations.[22, 23] The energies of the $^3$E′ and $^1$E′ states are too low in TDDFT due to their partial CT character (*vide infra*). While the energies of these higher states can possibly be improved by the use of range-separated functionals in TDDFT calculations, the correct ordering of the $S_1$ and $T_1$ states cannot be obtained without at least approximate inclusion of double excitations.[22, 29]

The highest occupied Hartree-Fock molecular orbitals of $\pi$ character (19 a″ - 22 a″ in $C_s$ symmetry) and n character (104 a′ - 106 a′ in $C_s$ symmetry) and the LUMO (23 a″ in $C_s$ symmetry) involved in the excitation of the four lowest singlet and triplet states of HAP-3MF are displayed in Fig. S1 in the supporting information. While the three highest occupied n orbitals are entirely localized on the Hz core, the four highest occupied $\pi$ orbitals are linear combinations of the HOMO of the Hz core and the highest occupied orbitals of the fluorotoluene substituents. Nevertheless, the electronic excitation spectrum of HAP-3MF exhibits the characteristic features of the excitation spectrum of Hz. The mixing of the $\pi$ orbitals of the Hz core with the $\pi$ orbitals of the substituents lowers the energies of the $^1$E′ and $^3$E′ excited states, while the energies of the low-lying $^1$A′ and $^3$A′ states are little affected by the substituents, as was previously observed for TAHz.[23]

The coefficients of the leading configurations in the ADC(2) wave functions of the excited states are also given in Table 1. It is seen that the wave functions of the $S_1$ and $T_1$ states are represented by excitations from two $\pi$ orbitals which are delocalized over the entire molecular system (19 a″, 20 a″) to the $\pi^*$ orbital (23 a″) which is localized on the Hz core. Moreover, the components of the 19 a″ and 20 a″ orbitals on the Hz core are mainly localized on the peripheral N-atoms, while the LUMO on the Hz core is localized on the peripheral C-atoms and the central N-atom. As a result of these atomic localization patterns, the exchange integral for the HOMO-LUMO transition is exceptionally small. Diagonalizing the electronic transition density matrices for the $S_1$ and $T_1$ states results in the natural transition orbitals (NTOs)[48] which are displayed in Fig. S2 in the supporting information. In this representation, the $S_0$-$S_1$ and $S_0$-$T_1$ transitions are strongly dominated by a single-electron excitation involving orbitals which are nearly exclusively localized on the Hz core.

It should be noted that the coefficient of the main configuration of the $S_1$ state (0.75) is slightly smaller than the corresponding coefficient of the $T_1$ state (0.77), see Table 1. This difference reflects the higher admixture of double excitations in the $S_1$ wave function which lowers the $S_1$ energy more than the $T_1$ energy, resulting in a negative $\Delta_{ST}$. This effect is also seen in the NTO representation. The natural orbital occupation numbers are 0.951 for the $S_1$ state and 0.960 for the $T_1$ state, see Fig. S2 in the supporting information.

It should also be noted that the lowest singlet and triplet $\pi\pi^*$ excited states of E symmetry exhibit a certain amount of CT character: electronic charge is transferred from the ligands to the Hz core, see the configurations in Table 1 and the Hartree-Fock orbitals in Fig. S1 in the supporting information. The CT character is reflected by substantial dipole moments of the components of the spatially degenerate states (see Table 1).

Geometry optimization of the $S_1$ and $T_1$ excited states of HAP-3MF with $C_{3h}$ symmetry constraint (technically, the ADC(2) calculations are performed in $C_s$ symmetry) results in vibrational stabilization energies of about 0.10 eV with respect to the VEEs for both $S_1$ and $T_1$ states, leaving the singlet-triplet energy gap unaffected. The NTOs and natural occupation numbers at the excited-state equilibrium geometries are very similar to their values at the ground-state equilibrium geometry (see data given in Fig. S3 in the supporting information).

Relaxation of all symmetry constraints in the excited-state geometry optimization results in a $C_{3h} \rightarrow C_3$ symmetry lowering due to a weak out-of-plane displacement of the central nitrogen atom. This stabilizes the $S_1$ and $T_1$ states by merely 0.04 eV, see data in Fig. S3 in the supporting information, and thus has little effect on the absorption spectrum. For the vertical emission spectra (fluorescence and phosphorescence, respectively), the effect of out-of-plane deformation is more pronounced (ca. 0.3 eV, see data in Fig. S3 in the supporting information). Overall, HAP-3MF is found to be rather rigid upon excitation to the $S_1$ and $T_1$ states and effects of excited-state geometry distortion play a minor role for its dynamics and spectroscopy, which agrees with the findings for TAHz.

HAP-3TPA-t exhibits $D_3$ symmetry in the electronic ground state with a planar tri-phenyl-Hz core and propeller-like twisted diphenylamino groups. The ADC(2) calculations have been performed in $C_2$ symmetry (choosing the symmetry axis along one of the C-C bonds connecting the diphenylaniline ligands with the Hz core). The VEEs of the three lowest singlet and triplet states of HAP-3TPA-t including oscillator strengths and dipole moments are listed in Table 2.

In contrast to HAP-3MF, the $^1n\pi^*$ states are located above the bright $^1\pi\pi^*$ state and therefore are not included in the Table. Notably, the ordering of the $^3A$ and $^3E$ states in HAP-3TPA is different from the ordering in HAP-3MF. The energy of the triplet partner of the $S_1$ state ($^3A_1$) is above the energy of the triplet partner ($^3E$) of the bright state $^1E$ state. While the $^1A_1$ - $^3A_1$ gap is -0.24 eV, the $^1A_1$ and $^3E$ VEEs are essentially degenerate at the ADC(2)/cc-pVDZ level, see Table 2. The interchange of the energies of the $^3A_1$ and $^3E$ states is the consequence of the conjugation in HAP-3TPA-t, which lowers the energies of the $^1E$ and $^3E$ states. While the energies of the $^1A$ and $^3A$ states are inverted (singlet below triplet), the energies of the $^1E$ and $^3E$ states are not inverted. These factors contribute to the near degeneracy of the $^1A$ and $^3E$ states.

*Table 2. Vertical excitation energies (ΔE), oscillator strengths (f), dipole moments (μ), and leading electronic configurations of the lowest excited states of HAP-3TPA-t determined with the ADC(2)/cc-pVDZ method at the MP2/ cc-pVDZ optimized equilibrium geometry of the electronic ground state. ADC(2)/def-SV(P) excitation energies computed at the MP2/def-SV(P) optimized equilibrium geometry of the electronic ground state are given in parentheses.*

| State | ΔE/eV | f | μ/Debye | Electronic Configuration |
|---|---|---|---|---|
| $S_0$ | 0.0 | - | 0.0 | $(121a)^2(115b)^2$ |
| $^1A_1(\pi\pi^*)$ | 2.59 (2.57) | 0.0 | 0.0 | 0.89(120a-116b)-0.31(117a-116b) |
| $^3E(\pi\pi^*)$ | 2.59 (2.62) | - | ±9.31 | 0.80(115b-116b)+0.32(114b-117b) 0.80(121a-116b)-0.32(114b-122a) |
| $^3A_2(\pi\pi^*)$ | 2.63 (2.67) | - | 0.0 | 0.76(114b-116b)+0.37(115b-117b)-0.37(121a-122a) |
| $^1E(\pi\pi^*)$ | 2.71 (2.75) | 1.946 | ±10.78 | 0.86(115b-116b)+0.27(114b-117b) 0.86(121a-116b)-0.27(114b-122a) |
| $^3A_1(\pi\pi^*)$ | 2.83 (2.84) | - | 0.0 | 0.91(120a-116b)-0.31(117a-116b) |
| $^1A_2(\pi\pi^*)$ | --- (2.96) | 0.0 | 0.0 | 0.86(114b-116b)+0.28(115b-117b)-0.27(121a-122a) |

The five highest occupied molecular π orbitals and the three lowest unoccupied orbitals of π* character are involved in the lowest electronic excitations of HAP-3TPA-t. The coefficients of the main configurations are included in Table 2. The Hartree-Fock molecular orbitals are displayed in Fig. S4 in the supporting information. Generally, the electronic wave functions comprise two electronic configurations involving different π orbitals which are linear combinations of orbitals of the Hz core and of the diphenylaniline substituents. The analysis of the electronic wave functions is more transparent in the NTO representation. The six NTOs involved in the lowest electronic states in Table 2 and their occupation numbers are given in Fig. S5 in the supporting information. In the NTO representation, it becomes clear that the excited states of $A_1$ symmetry are electronic excitations of the Hz core, while the excited states of E symmetry involve significant CT from the ligands to the Hz core. This CT character is also reflected by their substantial dipole moments (Table 2).

The NTO occupation numbers of the $^1A_1$ and $^3A_1$ electronic states of HAP-3TPA-t in Fig. S5 are similar to those of the $^1A$ and $^3A$ electronic states of HAP-3MF in Fig. S2. The higher admixture of double excitations in the $S_1$ state in comparison with the $T_1$ state stabilizes the $S_1$ state relative to the $T_1$ state. For the $^1E$ and $^3E$ states, the opposite behavior is observed. This explains the unusually low energy of the $^3E$ state (see Table 2 and data in Fig. S5).

The VEEs of HAP-3TPA-t obtained with the TDDFT/B3LYP method are given in Table S2 in the supporting information. While the energy of the $^1A_1(\pi\pi^*)$ state is qualitatively correctly predicted by TDDFT, the energies of the $^3E(\pi\pi^*)$ and $^3A_1(\pi\pi^*)$ states are grossly underestimated. Instead of the degeneracy of $^1A_1(\pi\pi^*)$ and $^3E(\pi\pi^*)$ predicted by ADC(2), TDDFT predicts the $^3E(\pi\pi^*)$ state 0.50 eV below the $^1A_1(\pi\pi^*)$ state. Since the CT character of the $^1E(\pi\pi^*)$ and $^3E(\pi\pi^*)$ states is more pronounced in HAP-2TPA-t than in HAP-3MF, the errors of TDDFT are larger for the former. Table S2 also gives the TDDFT VEEs of the full (not truncated) HAP-3TPA in comparison with the VEEs of HAP-3TPA-t. For the estimation of the effects of the tert-butyl groups TDDFT presumably is reliable, because the non-aromatic butyl groups are not involved in intramolecular electron transfer. It can be seen that the tert-butyl groups lower the VEEs by 0.1 eV or less and in a similar manner for singlet and triplet states.

Since the $^3E$ state is orbitally degenerate, it may exhibit a Jahn-Teller (JT) effect. The JT coupling lowers the energy of one component of the $^3E$ state by the JT stabilization energy, which is expected to remove the quasi-degeneracy of the vertical $^3E$ ($T_1$) and $^1A$ ($S_1$) states and the lower the energy minimum of the $T_1$ state below the energy minimum of the $S_1$ state. The optimization of stationary points on the potential-energy surface of the $^3E$ state was possible with the constraint of $C_2$ symmetry. This symmetry conserves the most important features of the $D_3$ symmetry of HAP-3TPA in the electronic ground state, such as the propeller-like conformation of the diphenylamino groups. In $C_2$ symmetry, the $S_1$ state ($^1A_1$ in $D_3$ symmetry) transforms as $^1B$, while the two component of the $^3E$ state transform as $^3A$ and $^3B$.

The adiabatic excitation energies of the $^1B$ state and the two components of the $^3E$ state are listed in Fig. S6 in the supporting information. The vibrational stabilization energies of the nondegenerate $^1B$ state and the degenerate $^3E$ state are 0.09 eV and 0.10 eV, respectively. Thus the degeneracy of the VEEs of the $^1B$ and $^3E$ states is lifted by only 0.01 eV in favour of the latter. While the vibrational stabilization energy of the $^3E$ state arises from the JT effect which involves in-plane modes and thus conserves the planarity of the Hz core, the nondegenerate $^1B$ excited state is stabilized by out of-plane deformation of the central N-atom of the Hz core. This effect is not included when the $^1B$ energy is optimized with $C_2$ symmetry constraint. Unfortunately, a fully unconstrained ($C_1$ symmetry) geometry optimization of the $^1B$ excited state of HAP-3TPA-t is not possible due to the size of the system. The effect of the out-of-plane symmetry breaking in the $^1B$ state of HAP-3TPA-t can qualitatively be estimated by comparison with smaller Hz-based systems for which geometry optimizations of the $S_1$ state and the $^3E$ state are possible. ADC(2)/cc-pVDZ calculations performed for Hz and tri-phenyl-Hz predict a vibrational stabilization energy of the $S_1$ state by about 0.04 eV, while the energy of the $^3E$ state is essentially unaffected by the $C_2 \rightarrow C_1$ symmetry lowering.

A schematic illustration of the potential-energy surfaces of the lowest singlet and triplet states of HAP-3TPA-t is shown in Fig. 2. The solid levels and bold-face numbers denote the adiabatic excitation energies of the $S_1$ state (blue) and of the $T_1$ state (red). The dashed levels and roman numbers denote the excitation energy of a given state computed at the equilibrium geometry of the other excited electronic state. Thus the $^3E$ energy is 0.16 eV above the $^1A$ energy at the minimum of the latter and the $^1A$ energy is 0.16 eV above the $^3E$ energy at the minimum of the latter. The energy difference between the two stationary points of the $^3E$ state (0.06 eV) represents the barrier for JT pseudo-rotation on the PE surface of the $^3E$ state. Italic numbers denote the vertical emission energies from the respective excited-state stationary points.

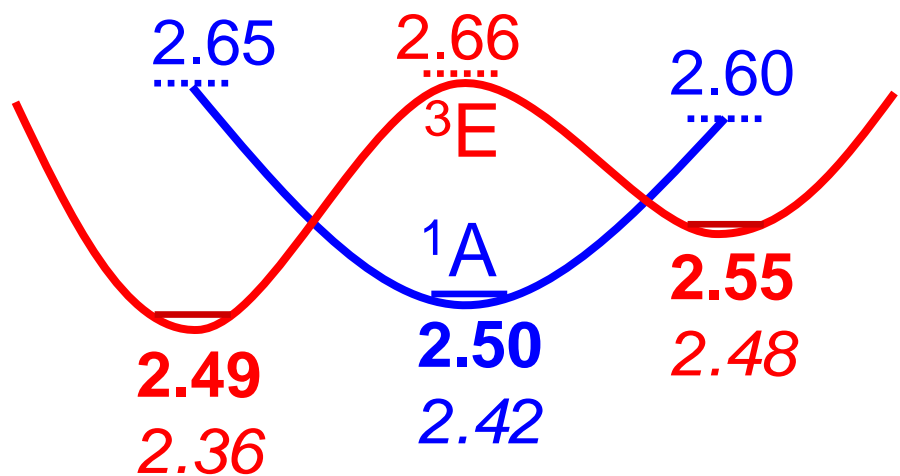

*Figure 2. Schematic illustration of the potential-energy surfaces of the lowest singlet and triplet excited states of HAP-3TPA-t. Solid levels and bold-face numbers denote the adiabatic excitation energies of the $S_1$ state (blue) and the $T_1$ state (red). Dashed levels and roman numbers denote the excitation energy of a given state computed at the equilibrium geometry of the other excited electronic state. The second minimum of the $^3E$ energy function represents the barrier for pseudo-rotation of the JT effect in the $^3E$ state. Italic numbers denote the vertical emission energies from the respective excited-state minima.*

The UV-vis absorption and luminescence spectra of HAP-3MF reported by Li et al.[33] are essentially identical with those of TAHz reported by Rabe et al.[26] These absorption spectra exhibit two weak transitions to $^1n\pi^*$ states near the onset of the strong absorption of the $^1E'$ state near 350 nm. The photoluminescence (PL) spectra extend from about 450 nm to about 650 nm and exhibit weak signatures of a vibronic structure.[26, 33, 34]

Based on TDDFT calculations and the dark character of the $S_1$ state, Li et al. assigned the $S_1$ and $T_1$ states of HAP-3MF as $^1n\pi^*$ and $^3n\pi^*$ states, respectively. As discussed above, these assignments are ruled out by the present more accurate ab initio calculations. The $S_1$ and $T_1$ states of HAP-3MF and TAHz rather are $\pi\pi^*$ states, see Table 1 and ref.[26]. The vanishing dipole transition moment of the $S_1(\pi\pi^*)$ state is a characteristic feature of Hz and $C_3$-symmetric derivatives of Hz. In trigonal symmetry, the $S_1 - S_0$ transition is dipole forbidden.

The peak of the fluorescence spectrum of HAP-3MF is observed at 525 nm (2.36 eV).[33] The calculated vertical emission energy of the $S_1$ state in $C_3$ symmetry is somewhat lower (2.06 eV), see data in Fig.

S3 in the supporting information. Li et al.[33] report an exceptionally long fluorescence lifetime of 252 ns for HAP-3MF in toluene which agrees well with the fluorescence lifetime of 287 nm reported by Rabe et al. for TAHz in toluene.[26] Li et al. observed that HAP-3MF shows no phosphorescence in frozen solution at 77 K. This finding, which is unusual for TADF systems, is in agreement with the analysis of the luminescence of TAHz described by Ehrmaier et al.[23]

Li et al. noted that it was difficult to observe photoluminescence from HAP-3MF which could be interpreted as TADF.[33] The missing of a clear TADF signal and the lack of phosphorescence were explained by a strongly suppressed SO coupling between the $S_1$ and $T_1$ states according to the El-Sayed rules. In time-resolved EL, on the other hand, delayed components of the luminescence were clearly observed. The percentage of the delayed fluorescence was as high as 71%, in EL, while in PL it was at most a few percent.[33] These observation are consistent with an IST system exhibiting a low ISC rate from the higher-lying $T_1$ state to the lower-lying $S_1$ state.

Additional spectroscopic data and luminescence efficiency benchmarks were reported by Adachi and coworkers for an exiplex consisting of HAP-3MF and mCP.[34] The HAP-3MF:mCP exciplex exhibits a slightly red-shifted emission, but the PL and EL characteristics are very similar to those reported for HAP-3MF. The spectral distributions of the prompt and delayed emissions were found to be nearly identical, from which it was concluded that $\Delta_{ST}$ is very small. In the light of the present ab initio results, this interpretation of the luminescence spectra has to be revised. The ab initio calculations predict a negative $\Delta_{ST}$ of about - 0.25 eV for both HAP-3MF and the isoelectronic TAHz. For the latter, the singlet-triplet inversion has been spectroscopically confirmed.[23] If all the luminescence originates from the origin of the $S_1$ state, the spectra of the prompt and delayed emissions obviously are identical. The transient lifetime of the prompt emission of HAP-3MF (184 ns) is typical for the fluorescence of the dark $S_1$ state of the Hz core, as discussed above. The transient lifetime of the delayed emission (2 - 6 μs), on the other hand, reflects the time scale of ISC from the higher-lying $T_1$ state to the lower-lying $S_1$ state. Since the ISC rate as well as the fluorescence rate are slow compared with typical intra-state vibrational relaxation (IVR) rates ($\approx$ 1 - 10 ps), the delayed emission originates from the origin of the $S_1$ state and therefore its spectral distribution is identical with that of the prompt emission. The observation of identical PL decay curves in air and under vacuum[34] is another indication that the luminescence originates from the $S_1$ state and not from a long-lived $T_1$ state. Since no phosphorescence is observed in HAP-3MF, it is not possible to determine $\Delta_{ST}$ directly from the luminescence spectra.

In 2013, Adachi and coworkers reported results for a highly efficient OLED using HAP-3TPA as emitter.[35] This device showed one of the best performances achieved with purely organic OLEDs at that time. Adachi and coworkers also reported detailed data on the PL and EL of HAP-3TPA. In contrast to HAP-3MF, phosphorescence of HAP-3TPA could be detected at 77 K. In oxygen-free toluene, a very high PL QE of 95% was observed and this high QE was maintained in a solid film.

Moreover, it is eye-catching that the spectra of the prompt and delayed emissions are nearly identical.[35]

The calculated VEE of the bright $S_2(\pi\pi^*)$ state of HAP-3TPA-t is 2.71 eV, which is in very good agreement with the observed absorption maximum of HAP-3TPA at 2.61 eV (475 nm).[35] Since this state is of partial CT character and therefore polar, it is expected that its VEE is redshifted in condensed-phase environments. The calculated vertical emission energy of the $S_1$ state of HAP-3TPA-t is 2.42 eV (513 nm) (see data in Fig. S6 in the supporting information), which compares well with the observed maximum of the prompt emission of HAP-3TPA at ≈ 2.30 eV (535 nm) in hexane at room temperature.[35] The radiative decay rate was found to be strongly temperature dependent, increasing from $2.9 \times 10^7$ s$^{-1}$ at 77 K to $1.4 \times 10^8$ s$^{-1}$ at 300 K. The QE of the prompt luminescence was found to increase from 19% to 85% as the temperature increased from 77 K to 300 K, while the QE of the delayed component was found to decrease from 57% to 6% over this temperature range.[35]

A strong temperature dependence of the fluorescence rate is expected to be a general phenomenon in trigonal Hz-based chromophores. The $S_1$ - $S_0$ transition dipole moment is inherently small in Hz due to the characteristic non-overlapping character of HOMO and LUMO, as discussed above. The $S_1$ - $S_0$ radiative transition is in addition symmetry forbidden in the $D_{3h}$ and $C_{3h}$ point groups. As the temperature rises, low-frequency non-totally symmetric vibrational levels become populated which induce a non-zero transition dipole moment by vibronic intensity borrowing from higher-lying bright electronic states. This effect can explain the pronounced increase of the fluorescence rate with temperature.

The fact that the increase of the prompt fluorescence is accompanied by a decrease of the delayed fluorescence provides information on the communication of the $S_1$ and $T_1$ vibronic manifolds via ISC/RISC. When the $S_1$ state is depopulated more rapidly by fluorescence at higher temperatures, population appears to flow from the $T_1$ state to the $S_1$ state, resulting in a decrease of the delayed luminescence. This implies a $T_1 \rightarrow S_1$ ISC/RISC rate which is of the same order of magnitude as the fluorescence rate of the $S_1$ state (≈ $10 \times 10^7$ s$^{-1}$). If the $T_1 \rightarrow S_1$ ISC/RISC rate were significantly smaller than the fluorescence rate of the $S_1$ state, the QE of the delayed fluorescence would not decrease with the same temperature gradient as the increase of the QE of the prompt fluorescence. This peculiar behaviour of the prompt and delayed QEs indicates that the populations of the vibronic levels of the $S_1$ and $T_1$ states are in quasi-equilibrium, which is to be expected if the minima of the $S_1$ and $T_1$ states are essentially isoenergetic. While the $S_1$ and $T_1$ states of HAP-3TPA are both $\pi\pi^*$ excitations, their orbital compositions are different, which may result in a larger SO coupling element than for the case of $S_1$ and $T_1$ states of the same orbital composition (as in HAP-3MF).[16, 49]

Another distinctive feature of HAP-3TPA is the strong dominance of the prompt luminescence component in PL ($\Phi_p/\Phi_d$ = 14.3), while in EL the prompt component is somewhat smaller than the delayed component ($\Phi_p/\Phi_d$ = 0.63).[35] The prompt PL is fed by ultrafast (≈ 100 fs) internal conversion from the bright $^1\pi\pi^*$ state to the S$_1$ state. At 300 K, the relaxed S$_1$ ($^1$A) state is depopulated by a fluorescence rate which may be higher than the ISC/RISC rate which equilibrates the S$_1$ and T$_1$ vibronic level populations. This can explain the very small delayed component in PL. In EL, on the other hand, the S$_1$ state is populated with a probability of only 25%. This population gives rise to the prompt fluorescence. The T$_1$ state is populated with a higher probability (75%) and gives rise to delayed luminescence. The observed $\Phi_p/\Phi_d$ of about 0.6 is larger than the statistical ratio of 1:3 of the singlet/triplet populations, which indicates that part of the triplet population undergoes RISC to the S$_1$ population and thus contributes to fluorescence within the chosen time window of $\approx$ 0.2 ms.[35] These numbers imply rather efficient equilibration of the singlet and triplet populations in HAP-3TPA, which is consistent with quasi-degenerate S$_1$ and T$_1$ states of different orbital character and the temperature dependence of the QEs discussed above.

The pronounced difference of the $\Phi_p/\Phi_d$ ratio for PL and EL in HAP-3TPA can be qualitatively rationalized as an effect of nuclear geometry by inspection of the qualitative potential-energy scheme in Fig. 2. In PL, the system is prepared, after ultrafast nonadiabatic relaxation from the optically prepared excited singlet state, in the vicinity of the equilibrium geometry of the $^1$A (S$_1$) state. In EL, on the other hand, the system is prepared in the vicinity of the equilibrium geometry of the JT-distorted $^3$E state. Since the fluorescence rate is strongly geometry dependent, as discussed above, and the ISC/RISC rate also may be geometry dependent, it is plausible that the $\Phi_p/\Phi_d$ ratio is very different in the two cases.

At the current level of computational accuracy, HAP-3TPA-t is predicted to be a TADF system with an exceptionally small positive adiabatic $\Delta_{ST}$ ($\approx$ 0.01 eV). HAP-3TPA thus appears to be a chromophore at the borderline of TADF and IST systems, which is compatible with the spectroscopic data of Adachi and coworkers.[35]

The example of HAP-3TPA-t illustration that in general it is not sufficient to compute just the vertical excitation energies of the S$_1$ and T$_1$ electronic states for the characterization of the mechanistic functionality of an OLED emitter. It is in general indispensable to explore the adiabatic energy surfaces of the relevant excited states (possibly also S$_n$ and T$_n$ states with n > 1) as well as their couplings in the vicinity of the Franck-Condon zone.[50] The example of HAP-3TPA-t also shows that rather accurate electronic-structure methods are necessary for meaningful computational screenings of TADF or IST emitters.

According to ab initio ADC(2) calculations, HAP-3MF is an example of an IST emitter with a negative vertical and adiabatic S$_1$-T$_1$ gap of about - 0.25 eV. HAP-3MF is isoelectronic with TAHz and exhibits very similar spectroscopic properties. HAP-3MF was implemented by Adachi and coworkers in an OLED device which showed a high maximum EQE of 11.3% and a low roll-off characteristic.[34] HAP-3MF presumably is the *first IST emitter* which was tested and characterized in an OLED device. The PL properties of HAP-3MF measured by Adachi and coworkers were originally interpreted in terms of TADF, assuming a small positive $\Delta_{ST}$.[33] However, several aspects of the spectroscopic data, such as the lack of phosphorescence at 77 K, a high QE of prompt fluorescence in PL, and a high QE of delayed luminescence in EL, are features which are expected for IST emitters.

HAT-3MF is an exemplar of Hz derivatives with "small" substituents. The spectroscopic properties of such derivatives are determined by the Hz core. The fluorotoluene substituents in HAP-3MF lower the energy and increase the oscillator strength of the bright $^1\pi\pi^*$ state, but have little impact on the spectroscopy and photophysical dynamics of the $S_1$ and $T_1$ states. It should be noted that the substituents are also essential for the photochemical stability of the Hz chromophore. As is well known, unsubstituted Hz hydrolyses spontaneously in the presence of traces of water,[51] forming a thermodynamically stable colorless photohydrate.[52] Recent advances in the synthesis of Hz derivatives with easily exchangeable groups[53] should pave the way for future systematic design studies leading to Hz-based OLED emitters with optimized quantum efficiencies and operando stabilities. The synthetic efforts could be guided by ab initio wave-function-based electronic structure calculations which allow accurate predictions of the sign and magnitude of the energy gap of the lowest singlet and triplet states.

HAP-3TPA, on the other hand, is an exemplar of Hz derivatives with large conjugated aromatic side groups, which have a more significant effect on the spectroscopy, as is revealed by the present ADC(2) calculations. In addition to the $^1$A' and $^3$A' states, which primarily arise from the HOMO-LUMO excitation of the Hz core, the triplet partner ($^3$E'($\pi\pi^*$)) of the bright $^1$E($\pi\pi^*$) state becomes relevant. The $^1$E($\pi\pi^*$) and $^3$E($\pi\pi^*$) states involve orbital excitations which are of mixed Hz/diphenylaniline character. Other than in HAP-3MF, the VEE of the lowest triplet state ($^3$E($\pi\pi^*$)) is predicted to be degenerate with the VEE of the $S_1$ state in HAP-3TPA-t. Both the $S_1$ state and the $T_1$ state exhibit moderate vibrational stabilization energies, the former being stabilized by out-of-plane deformation of the Hz core, the latter by the lifting of its degeneracy by the JT effect. The marginally larger vibrational stabilization energy of the $S_1$ state results in a predicted adiabatic $\Delta_{ST}$ of + 0.01 eV at the ADC(2) level. Most likely, HAP-3TPA is a chromophore with quasi-degenerate minima of the $S_1$ and $T_1$ states.

The observed PL and EL spectra of HAP-3TPA are compatible with this computational prediction. In contrast to HAP-3MF, weak phosphorescence can be observed at 77 K, which indicates that the lowest vibronic levels of the SO-coupled $S_1$-$T_1$ system must exhibit some triplet character. The strong dominance of prompt luminescence over delayed luminescence after photoexcitation at 300 K reveals that the fluorescence rate is higher than the ISC rate. On the other hand, the weak dominance of delayed luminescence over prompt luminescence in EL indicates that the $T_1 \rightarrow S_1$ ISC rate can effectively compete with the phosphoresce rate in HAP-3TPA. The unusual temperature dependence of the PL QE can be explained by a quasi-degenerate singlet-triplet system with vibronic wave functions which are strongly mixed by SO coupling.

The calculated dipole moment of the $^3$E state of HAP-3TPA-t is relatively large (9.3 Debye), whereas the dipole moment of the S$_1$ state is zero by symmetry. The amount of mixing of the singlet/triplet vibronic levels and the luminescence properties of HAP-3TPA therefore should be rather sensitive to the environment. Future investigations of the PL and EL of HAP-3TPA in different environments could shed additional light on the unusual singlet-triplet mixing dynamics in this intriguing chromophore.

**Acknowledgments**

This work was funded by the National Science Center of Poland, Grant No. 2020/39/B/ST4/01723.

**Supporting Information Available:**

Vertical excitation energies of HAP-3MF and HAP-3TPA-t obtained with the TDDFT/B3LYP method; highest occupied and lowest unoccupied Hartree-Fock molecular orbitals of HAP-3MF and HAP-3TPA-t; natural transition orbitals and occupation numbers of HAP-3MF and HAP-3TPA-t; adiabatic excitation energies and vertical emission energies of HAP-3MF and HAP-3TPA-t.

**Supporting Information**

**Are Heptazine-Based Organic Light-Emitting Diode Chromophores Thermally Activated Delayed Fluorescence or Inverted Singlet-Triplet Systems?**

Andrzej L. Sobolewski[1] and Wolfgang Domcke[2]

[1] Institute of Physics, Polish Academy of Sciences, Warsaw, Poland

[2] Department of Chemistry, Technical University of Munich, Garching, Germany

*Table S1. Vertical excitation energies (ΔE) and oscillator strengths (f) of the lowest excited states of HAP-3MF, determined with the TDDFT/B3LYP/cc-pVDZ method at the DFT-D3/B3LYP/cc-pVDZ equilibrium geometry of the electronic ground state.*

| state | ΔE/eV | f |
|---|---|---|
| $^3$A'($\pi\pi^*$) | 2.64 | - |
| $^1$A'($\pi\pi^*$) | 2.81 | 0.0 |
| $^3$E'($\pi\pi^*$) | 2.94 | - |
| $^3$A''($n\pi^*$) | 3.45 | - |
| $^1$E'($\pi\pi^*$) | 3.52 | 0.880 |
| $^3$E''($n\pi^*$) | 3.54 | - |
| $^1$A''($n\pi^*$) | 3.55 | 0.0 |
| $^1$E''($n\pi^*$) | 3.65 | 0.0 |

Table S2. Vertical excitation energies (ΔE) and oscillator strengths (f) of the lowest excited states of HAP-3TPA and truncated HAP-3TPA (HAP-3TPA-t), determined with the TDDFT/B3LYP/cc-pVDZ method at the DFT-D3/B3LYP/cc-pVDZ equilibrium geometry of the electronic ground state.

|  | HAP-3TPA | | HAP-3TPA-t | |
|---|---|---|---|---|
| state | ΔE/eV | f | ΔE/eV | f |
| $^3E(\pi\pi^*)$ | 2.15 | - | 2.22 | - |
| $^3A_1(\pi\pi^*)$ | 2.20 | - | 2.27 | - |
| $^1E(\pi\pi^*)$ | 2.43 | 1.607 | 2.52 | 1.530 |
| $^3A_2(\pi\pi^*)$ | 2.65 | - | 2.65 | - |
| $^1A_1(\pi\pi^*)$ | 2.65 | 0.0 | 2.74 | 0.0 |
| $^1A_2(\pi\pi^*)$ | 2.81 | 0.0 | 2.81 | 0.0 |

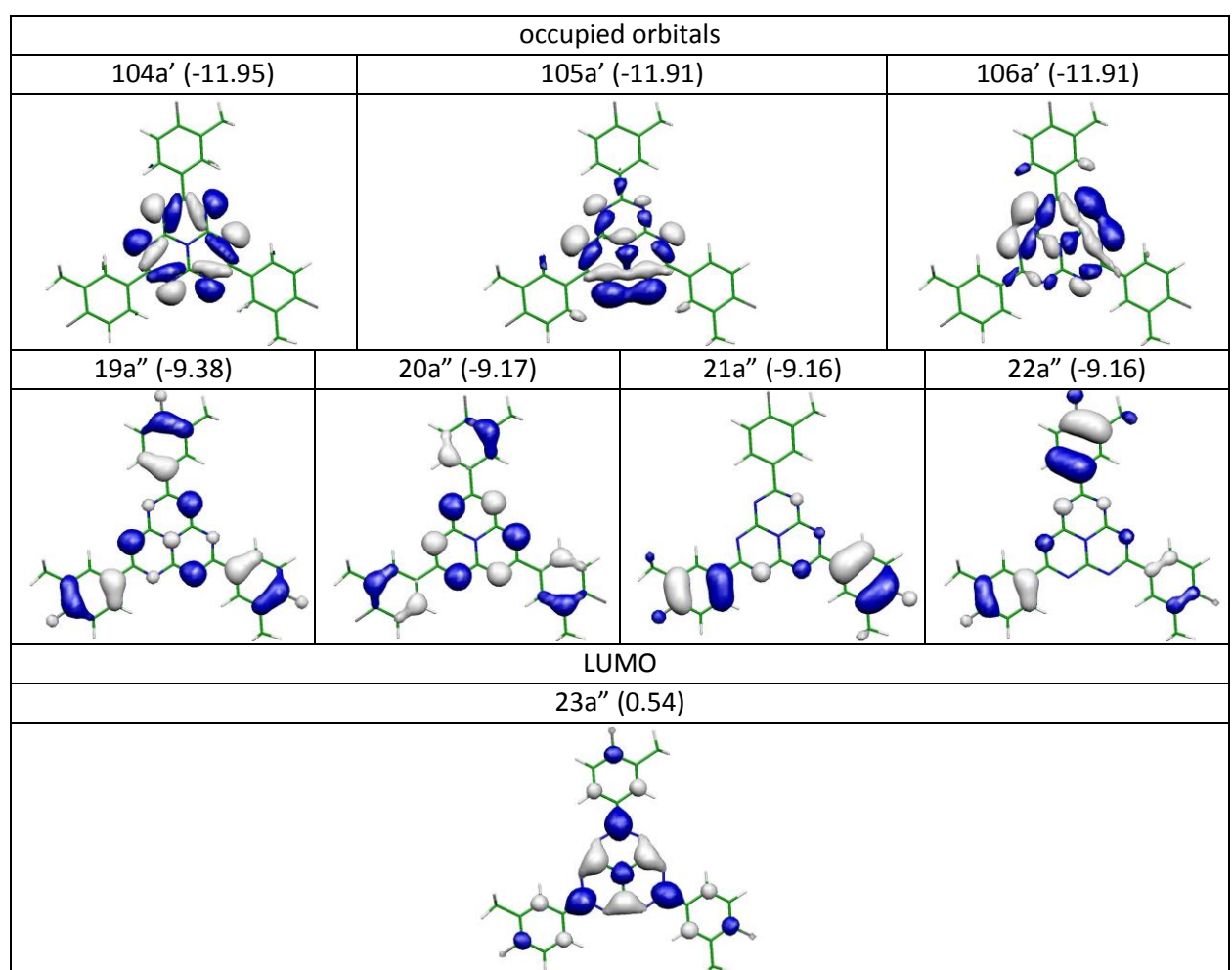

Figure S1. The highest occupied and lowest unoccupied Hartree-Fock orbitals involved in the lowest electronic excitations of HAP-3MF at the ground-state equilibrium geometry. The orbital energies (in eV) are shown in parentheses. Orbitals were plotted with an isosurface value of 0.03.

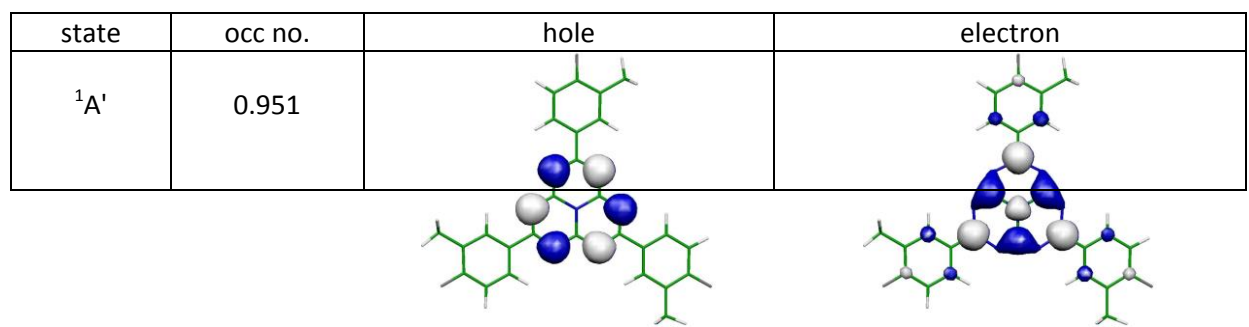

| state | occ no. | hole | electron |
|---|---|---|---|
| $^1A'$ | 0.951 | | |
| ³A' | 0.960 | | |

*Figure S2. Occupation numbers and natural transition orbitals involved in the excitation to the ¹A' and ³A' excited states of HAP-3MF at the ground-state equilibrium geometry. Orbitals were plotted with an isosurface value of 0.03.*

| state | $E_{ad}$/eV | $E_{em}$/eV | occ. no. | hole | electron |
|---|---|---|---|---|---|
| $C_{3h}$ symmetry | | | | | |
| ¹A' | 2.47 | 2.42 | 0.951 | 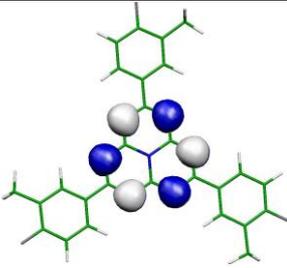 | 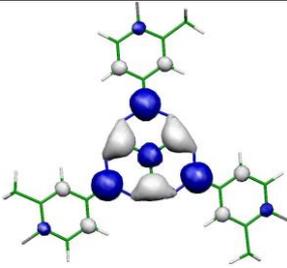 |
| ³A' | 2.72 | 2.67 | 0.974 | | |
| $\Delta_{ST}$ | -0.25 | -0.25 | | | |
| $C_3$ symmetry | | | | | |
| $S_1$ | 2.43 | 2.06 | 0.955 | 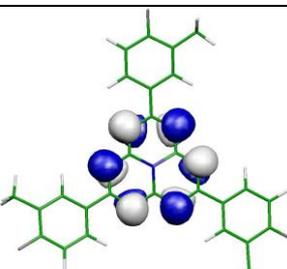 | 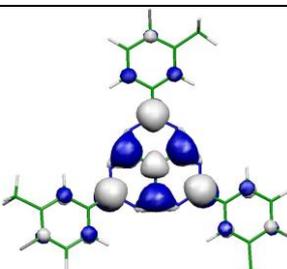 |
| $T_1$ | 2.69 | 2.36 | 0.957 | | |
| $\Delta_{ST}$ | -0.26 | -0.30 | | | |

*Figure S3. Adiabatic excitation energies ($E_{ad}$), vertical emission energies ($E_{em}$), natural occupation numbers, and natural transition orbitals of the lowest excited states of HAP-3MF, determined with the ADC(2)/cc-pVDZ method with the indicated symmetry constraints. Orbitals were plotted with an isosurface value of 0.03.*

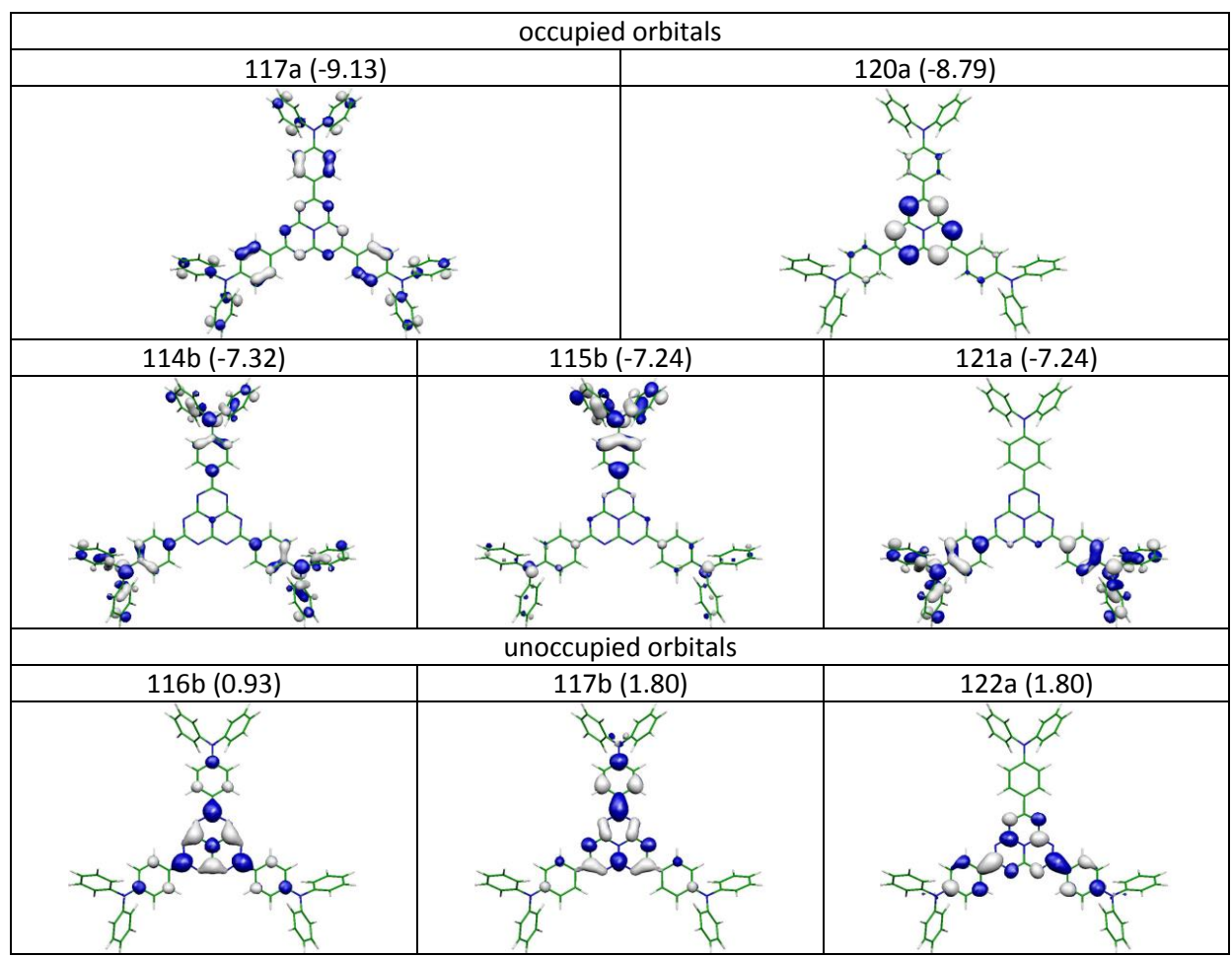

*Figure S4. The highest occupied and lowest unoccupied Hartree-Fock orbitals involved in the lowest electronic excitations of HAP-3TPA-t at the ground-state equilibrium geometry. The orbital energies (in eV) are shown in parentheses. Orbitals were plotted with an isosurface value of 0.03.*

| state | occ no. | hole | electron |
|---|---|---|---|
| $^1A_1$ | 0.949 | 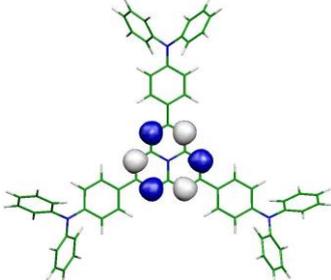 | 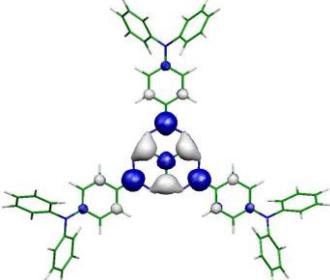 |
| $^3A_1$ | 0.957 | | |
| $^1E_x$ | 0.831 | 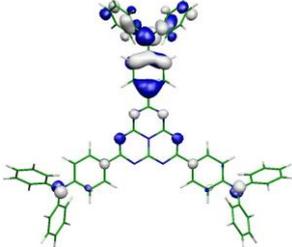 | 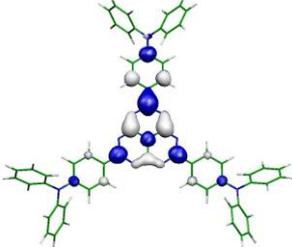 |
| $^3E_x$ | 0.770 | | |
| $^1E_y$ | 0.827 | 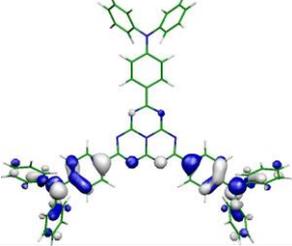 | 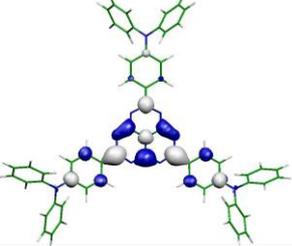 |
| $^3E_y$ | 0.759 | | |

*Figure S5. Occupation numbers and natural transition orbitals involved in the excitation to the lowest excited electronic states of HAP-3TPA-t at the ground-state equilibrium geometry. Orbitals were plotted with an isosurface value of 0.03.*

| state | $E_{ad}$/eV | $E_{em}$/eV | occ. no. | hole | electron |
|---|---|---|---|---|---|
| $^1$B | 2.50 | 2.42 | 0.950 | 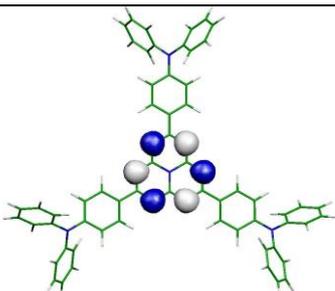 | 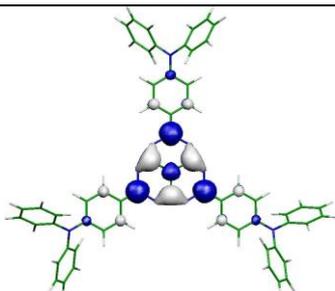 |
| $^3$E$_x$ | 2.49 | 2.36 | 0.947 | 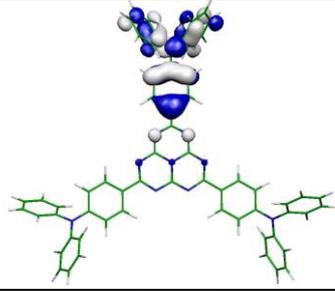 | 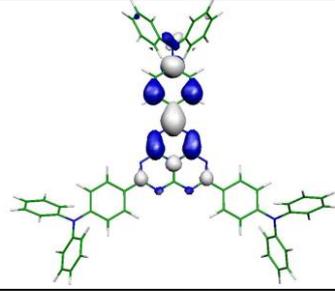 |
| $^3$E$_y$ | 2.55 | 2.48 | 0.768 | 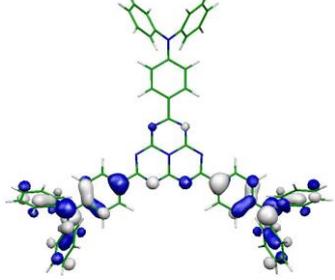 | 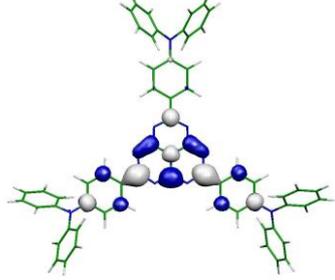 |

*Figure S6. Adiabatic excitation energies ($E_{ad}$) vertical emission energies ($E_{em}$), natural occupation numbers, and natural transition orbitals of the lowest excited states of HAP-3TPA-t determined with the ADC(2)/def-SV(P) method at the equilibrium geometry of a given excited state. Orbitals were plotted with an isosurface value of 0.03.*